\documentclass[pra, superscriptaddress, twocolumn, showpacs]{revtex4}

\usepackage{amsfonts}
\usepackage{amssymb}
\usepackage[dvips]{graphicx}
\usepackage{amsmath}
\usepackage{color}

\newcommand{\bra}[1]    {\langle #1\vert}
\newcommand{\ket}[1]    {\vert #1 \rangle}    
\newcommand{\e}         {\mathrm{e}}
\newcommand{\ii}        {\mathrm{i}}

\begin{document}

\title{Quantum frequency estimation with trapped ions and atoms}

\author{U.~Dorner}
\affiliation{Centre for Quantum Technologies, National University of Singapore, 3 Science Drive 2, Singapore 117543, Singapore}
\affiliation{Clarendon Laboratory, University of Oxford, Parks Road, Oxford OX1 3PU, United Kingdom}

\date{\today}


\begin{abstract} 
  We discuss strategies for quantum enhanced estimation of atomic
  transition frequencies with ions stored in Paul traps or neutral
  atoms trapped in optical lattices.  We show that only marginal
  quantum improvements can be achieved using standard Ramsey
  interferometry in the presence of collective dephasing, which is the
  major source of noise in relevant experimental setups.  We therefore
  analyze methods based on decoherence free subspaces and prove that
  quantum enhancement can readily be achieved even in the case of
  significantly imperfect state preparation and faulty detections.
\end{abstract}

\maketitle

The ultra-precise estimation of physical parameters is of great
importance for countless applications, such as atomic clocks,
gravitational wave detectors, laser gyroscopes or microscopy. Quantum
enhanced precision measurements have the potential to significantly
increase the precision of parameter estimation compared to classical
methods in such applications~\cite{glm04}. This is generally achieved
by preparing a quantum probe-state which has a higher sensitivity with
respect to the quantity to be estimated. Assuming that this probe
consists of $N$ non-interacting, identical subsystems (e.g. $N$
particles), the estimation uncertainty in many applications including
optical or atomic interferometry can then ideally be improved from the
standard quantum limit (SQL), which scales like $1/\sqrt{N}$, to the
Heisenberg limit, which scales like $1/N$~\cite{glm06,Boixo07}.
Endeavors to attain the Heisenberg limit (or at least to beat the SQL)
are made in many branches of physics including quantum
photonics~\cite{Dowling08} and atomic
physics~\cite{Leibfried04,Meyer01,Leroux10,Polzik10,Gross10,Riedel10}.  In atomic physics, the
measurement of atomic transition frequencies with Ramsey
interferometry has been established as an important tool, not only for
general spectroscopic purposes but also to determine frequency
standards on which atomic clocks are based on~\cite{Diddams04}.
Improvements of Ramsey interferometry via quantum effects are
therefore highly desirable. As in other quantum technologies like
quantum computing and communication, the biggest obstacle for the
realization of such a quantum interferometer is the presence of
unavoidable noise and imperfections.  A practical quantum sensor must
therefore use probe-states which are robust under realistic
circumstances, as well as preparation and detection schemes which can
be performed with high fidelity.

In this paper we analyze methods for quantum enhanced estimation of
atomic transition frequencies with Ramsey interferometry, and
generalizations thereof, which can improve the measurement uncertainty
to the Heisenberg limit in the presence of noise, and which tolerate
imperfect state preparation and detection. A scheme for quantum
enhanced Ramsey interferometry has been proposed some time
ago~\cite{Bollinger96}, but it has subsequently been shown that in the
presence of noise, in form of {\em uncorrelated} dephasing, the scheme
has only little or no advantage compared to its classical
counterpart~\cite{Huelga97}. However, recent experimental
breakthroughs with closely spaced particles, particularly ions stored
in linear Paul traps~\cite{Monz10,Roos06,Langer05,Kielpinski01}, show
that the major source of noise in these systems consists of {\em
  correlated} dephasing. Motivated by this insight we first analyze
conventional Ramsey interferometry and show that in the presence of
correlated dephasing hardly any quantum enhancement can be achieved.
We therefore discuss alternative methods which make use of decoherence
free subspaces~\cite{Lidar03} and show that they lead to quantum
enhanced precision even in the presence of significantly imperfect
state preparation and faulty detections. Our approach is mainly
motivated by recent experiments with trapped ions, but it can also be
applied to cold atoms stored in optical lattices~\cite{Jaksch05}. The
main body of this paper concisely summarizes our results. Details of
calculations can be found in the appendices.
\begin{figure}[b]
\centering\includegraphics[width=7.65cm]{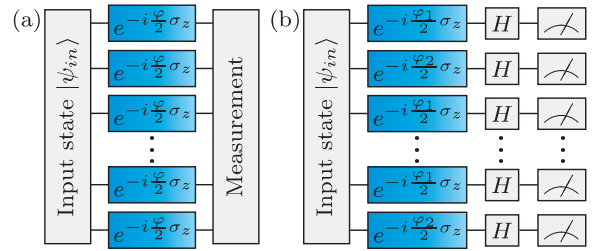}
\caption{%
  (a) Schematic Ramsey interferometer. (b) Generalized setup where
  atoms acquire different phases (see text).}
\label{fig1}
\end{figure}

We consider $N$ two-level atoms or ions, with internal states
$\ket{0}$ and $\ket{1}$, which are, e.g., stored in a linear Paul
trap. These atoms are prepared in an input state $\ket{\psi_{in}}$ and
undergo the process shown in Fig.~\ref{fig1}(a), i.e. each atom
accumulates a phase $\varphi$ during a time $t$ and is finally
measured. The process is repeated $\nu$ times and based on the
measurement outcomes the phase can be estimated.  For simplicity we
assume that the measurement and preparation times are much smaller
than $t$, such that the total time of the experiment is given by
$T=\nu t$.  It is our goal to make the uncertainty of the estimated
phase as small as possible for a given $T$ and $N$. 

In conventional Ramsey interferometry the input state is given by the
product state $\ket{\psi_{in}^{pro}}= [(\ket{0}+\ket{1})/\sqrt2
]^{\otimes N}$ which is prepared by a $\pi/2$-pulse using a laser with
frequency $\omega_L$ which is slightly detuned from the atomic
transition frequency $\omega$. Note that for simplicity in this paper
we identify $\pi/2$-pulses with Hadamard gates which has no effect on
the estimation uncertainty. Each atom then undergoes a free evolution
of duration $t$ before a second $\pi/2$-pulse (using the same laser)
and a measurement of the atomic state is performed. During the time
$t$ the atoms gather up a relative phase $\varphi=(\omega-\omega_L)t$
which can be estimated from the measurement data. If $\omega_L$ and
$t$ are known, we therefore obtain an estimate $\omega_{est}$ of the
frequency $\omega$ with an uncertainty given by~\cite{Braunstein94,Braunstein96}
\begin{equation}
\Delta\omega = 
\left\langle \left( \frac{\omega_{est}}{|\partial\langle \omega_{est}\rangle/\partial\omega|}-\omega  \right)^2  \right\rangle^{1/2}
\label{eq:uncertainty}
\end{equation}
which, for unbiased estimators, is simply the standard deviation. The
uncertainty, or precision, $\Delta\omega$ is bounded from below by the
(quantum) Cram\'er-Rao bound~\cite{Helstrom,Braunstein94,Braunstein96}
\begin{equation}
\Delta\omega \ge \frac{1}{\sqrt{\nu F}}\ge \frac{1}{\sqrt{\nu F_Q}} =   
\frac{1}{\sqrt{T F_Q/t}} \equiv \Delta\omega_{min},
\label{eq:crb}
\end{equation}
where $F$ is the Fisher information and $F_Q$ is the quantum Fisher
information (QFI). Expressions for $F$ and $F_Q$ can be found
in~\cite{Braunstein96,Demkowicz09} and Appendices~\ref{sec:AppII},
\ref{sec:AppIII}. The Fisher information depends on the state of the
system before the measurement and the measurement itself while the QFI
depends only on the state before the measurement.  The first bound in
Eq.~(\ref{eq:crb}) can be reached via maximum likelihood 
estimation for large $\nu$ (or $T$) and the second bound by an {\em
  optimal} measurement which always exists~\cite{Braunstein94}.

If we assume that our pure input state remains pure, a product state
$\ket{\psi_{in}^{pro}}$ as input then leads to the SQL precision
$\Delta\omega_{min} = 1/\sqrt{TtN}$, whereas an entangled
($N$-particle) Greenberger-Horne-Zeilinger (GHZ) state,
$\ket{\psi_{in}^{GHZ}}= ( \ket{0}^{\otimes N} + \ket{1}^{\otimes N} )
/\sqrt2$, improves the precision to $\Delta\omega_{min} =
1/\sqrt{Tt}N$, i.e. the Heisenberg limit~\cite{Bollinger96}.  However,
under realistic conditions, the pure input state will degrade into a
mixture due to unavoidable noise.  For the systems considered here the
dominant source of noise is dephasing caused by fluctuating (stray)
fields leading to random energy shifts of the atomic levels.  As shown
in~\cite{Huelga97,Shaji07}, the advantage of a GHZ state deteriorates
in case of {\em uncorrelated} dephasing, leading to exactly the same
optimal precision as a product state which has merely SQL
scaling. However, this is not the situation commonly encountered in
ion traps or atoms in optical lattices where particles are very
closely spaced. Here, the particles are subject to the same
fluctuations which leads to {\em correlated} dephasing such that the
time evolution of the system state $\rho$ is determined by 
\begin{equation}
\dot\rho = -\ii\frac{\delta}{2}[S_z,\rho] + \frac{\gamma}{2} 
\left( L\rho L -\frac{1}{2}L^2\rho - \frac{1}{2}\rho L^2 \right), 
\label{eq:dyn1}
\end{equation}
where $\delta=\omega-\omega_L$, $\gamma$ is a dephasing rate and
$L=S_z\equiv\sum_{j=1}^N\sigma_z^j$, where $\sigma_z^j$ is the Pauli
$z$-operator acting on atom $j$. The fact that Eq.~(\ref{eq:dyn1})
describes the dominant source of noise in the setups considered in
this paper was very clearly shown in a number of recent
experiments~\cite{Monz10,Roos06,Langer05,Kielpinski01}. Equation~(\ref{eq:dyn1})
can be derived via a Langevin-equation approach by assuming that the
atoms are subject to level shifts caused by the same fluctuating field
with a sharp time correlation function (see Appendix~\ref{sec:AppI}
for details). This is typically the case in experiments, e.g. in the
ion trap experiment~\cite{Monz10}, where the dominant source of noise
is caused by fluctuations of the homogeneous magnetic field which is required
to lift Zeeman degeneracies and which affects all ions in an equal manner.
\begin{figure}[t]
\centering\includegraphics[]{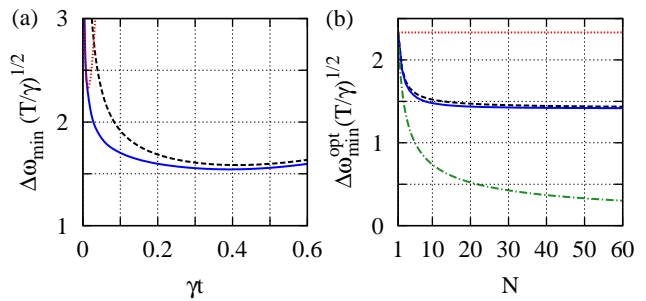} 
\caption{%
  (Color online) (a) Precision $\Delta\omega_{min}$ versus duration of
  the free evolution for $N=6$ atoms. The minima of the curves define
  $\Delta\omega_{min}^{opt}$. (b) Best possible precision
  $\Delta\omega_{min}^{opt}$ versus atom number $N$. In both figures
  the dotted (red) line corresponds to a GHZ state, the dashed (black)
  line to a product state and the solid (blue) line to the optimal
  precision. In (b) we also show the precision corresponding to a
  product/GHZ state undergoing uncorrelated dephasing (dashed-dotted,
  green line).}
\label{fig2}
\end{figure}

Suppose we consider states which are symmetric under particle
exchange, then we can use a Fock representation in which $S_z=n_0 -
n_1$, where $n_i\equiv a_i^\dagger a_i$, and $a_i^\dagger$ ($a_i$) are
bosonic creation (annihilation) operators of an atom in state
$\ket{i},\,i=0,1$. We can then rewrite Eq.~(\ref{eq:dyn1}) to obtain
\begin{equation}
\dot\rho = -\ii\delta[n_ 0,\rho] + 2\gamma\left [ n_ 0 \rho n_ 0
  -\frac{1}{2} n_ 0^2 \rho -\frac{1}{2} \rho n_ 0^2 \right], 
\label{eq:master0}
\end{equation}
and a symmetric, pure input-state has the form $\ket{\psi_{in}} =
\sum_{k=0}^N \alpha_k \ket{k,N-k}$, where $\ket{k,N-k}$ is a Fock
state with $k$ ($N-k$) atoms in state $\ket{0}$ ($\ket{1}$).
Equation~(\ref{eq:master0}) can be solved analytically [see
Eq.~(\ref{eq:rho})] which yields the system state immediately before
the measurement which can be used to calculate the QFI.  A GHZ state
is in this representation formally equivalent to a NOON-state known
from optical interferometry~\cite{Dowling08}, $\ket{\psi_{in}} =
(\ket{N,0}+\ket{0,N})/\sqrt2$. Using this state as initial state
leads, via the QFI, to the precision
\begin{equation}
\Delta\omega_{min} 
= \frac{1}{\sqrt{T t} N\e^{-\gamma N^2 t}}
\ge \sqrt{ \frac{2\e \gamma}{T}}\equiv \Delta\omega_{min}^{opt}.
\end{equation}
The quantity $\Delta\omega_{min}^{opt}$ is found by using an optimal
time $t_{opt}=1/2\gamma N^2$ for each experimental run. As can be
seen, $\Delta\omega_{min}^{opt}$ has no dependency on $N$ and
therefore there is no advantage using a GHZ state in the presence of
collective dephasing. The precision $\Delta\omega_{min}$ and
$\Delta\omega_{min}^{opt}$ corresponding to a product state can be
calculated numerically leading to $\Delta\omega_{min}^{opt} \approx
(\sqrt{2} + 0.87/N^{0.90})\sqrt{\gamma/T}$ which has no SQL scaling
and does not even approach zero for large $N$, but is still better
than the GHZ case (see Fig.~\ref{fig2}).

A decisive feature of collective dephasing is the existence of
decoherence free subspaces (DFSs)~\cite{Lidar03} which are given by
states such that $L\ket{\psi_{DFS}}=0$.  However, since $L=S_z$ in
Eq.~(\ref{eq:dyn1}) a highly robust DFS state would be stationary and
hence useless for frequency estimation. Ideally, one would therefore
use input states which lead to an optimal trade-off between gain in
precision and robustness which can be found by maximizing the QFI with
respect to all possible input states. In Appendix~\ref{sec:AppII} we
show that the maximal QFI can be attained by states which are
symmetric under particle exchange leading to a considerable
simplification of the optimization problem. Results are shown in
Fig.~\ref{fig2}. As can be seen the best possible state leads only to
a marginal improvement over a product state. For comparison, in
Fig.~\ref{fig2}(b) we also plot the precision corresponding to
$\ket{\psi_{in}^{pro}}$ which is subject to uncorrelated
dephasing~\cite{Huelga97}.  Evidently, correlated dephasing is
significantly more destructive than uncorrelated dephasing.

To make use of the coherence preserving features of DFSs we have to
alter the dynamics of the system such that the incoherent part of
Eq.~(\ref{eq:dyn1}) is zero and the coherent part is non-zero. To this
end we consider a scheme consisting of $N$ atoms ($N$ even), where
half of the atoms accumulate a phase $\varphi_1$ and the other half
$\varphi_2$. Such a scheme is realized in a system where, e.g., every
odd atom has a transition frequency $\omega_1$ and every even atom has
a transition frequency $\omega_2$ [see Fig.~\ref{fig1}(b)] and our
goal is to estimate the frequency difference
$\delta\equiv\omega_1-\omega_2$ (see Appendix~\ref{sec:DFS1}).  If fluctuating fields lead to the
same energy shift in both transitions the incoherent part of the
master equation~(\ref{eq:dyn1}) vanishes if a DFS state of the form
$\ket{\psi_{in}} = (\ket{0101\ldots01} +\ket{1010\ldots10})/\sqrt2$ is
used.  Via the QFI, we then obtain the bound for the precision
$\Delta\delta_{min} = 2/\sqrt{Tt}N$ which has Heisenberg scaling even
in the presence of correlated dephasing.  We should note here that in
general we can use arbitrary orderings of the atoms in
Fig.~\ref{fig1}(b). The input state then takes the form
\begin{equation}
\ket{\psi_{in}} = \frac{1}{\sqrt 2}( \ket{i_1,i_2,\ldots,i_N} +
\prod_{j=1}^N\sigma_x^j\ket{i_1,i_2,\ldots,i_N} ),
\end{equation} 
where $i_j=0,1$ and $\sum_{j=0}^N i_j = N/2$ (i.e. $i_j=0$ occurs as
many times as $i_j=1$) and atoms with $i_j=0$ ($i_j=1$) accumulate the
phase $\varphi_1$ ($\varphi_2$).  The described dynamics can be
obtained, e.g., by choosing the two transitions to be within the same
Zeeman manifold such that the difference of the magnetic quantum
numbers of each transition is equal, and thus a (weak) fluctuating
magnetic field leads to the same energy shifts. This was demonstrated
in a recent experiment with two ions in a linear Paul trap revealing a
significant increase in the coherence time~\cite{Roos06}. The same
ideas were furthered by an experimental study of non-perfect input
states~\cite{Chwalla07}. In both experiments, an additional electric quadrupole field was
used to obtain $\omega_1\ne\omega_2$ and from the measured frequency
difference the electric quadrupole moment was determined.

The above experiments also offers an alternative view of the fact that
Heisenberg scaling can be obtained in this setup. In
Ref.~\cite{Roos06} a `designer atom' was constructed consisting of two
physical atoms with two internal, logical states
$\ket{0}_L\equiv\ket{01}$ and $\ket{1}_L\equiv\ket{10}$ which are
decoherence free. A state of the form
$(\ket{0101\ldots01}+\ket{1010\ldots10})/\sqrt2$ is then equivalent to
a decoherence free GHZ state of $n=N/2$ designer atoms,
$\ket{\psi_{in}}_L=(\ket{00\ldots0}_L+\ket{11\ldots1}_L)/\sqrt2$. The
states $\ket{0}_L,\,\ket{1}_L$ accumulate a relative phase $\delta t$,
and so it is straightforward that $\ket{\psi_{in}}_L$ leads to a
sensitivity $\Delta\delta_{min}=1/\sqrt{Tt}n$.

We can also conceive a situation where the fluctuating field shifts
the transition frequency $\omega_1$ of half of the atoms and the
transition frequency $\omega_2$ of the other half by the same
magnitude but opposite sign. For the setup shown in
Fig.~\ref{fig1}(b), this means that we have to replace the noise
operator in Eq.~(\ref{eq:dyn1}) by $L = \sum_{j=1}^N (-1)^j\sigma_z^j$
and a GHZ state would be decoherence free. We can utilize this for
quantum enhanced precision measurements by performing a Ramsey-type
experiment but now with up to two lasers of frequency $\omega_{L1}$
and $\omega_{L2}$ such that half of the atoms accumulate a relative
phase $\varphi_1=(\omega_1 - \omega_{L1})t\equiv\delta_1t$ and the
other half $\varphi_2=(\omega_2 - \omega_{L2})t\equiv\delta_2t$.  The
Hamiltonian can then be written as (see Appendix~\ref{sec:DFS2})
\begin{equation}
H = \frac{1}{4}(\delta_1+\delta_2)S_z + \frac{1}{4}(\delta_2-\delta_1)L,
\end{equation}
where the second term vanishes if applied to a GHZ state. If the laser
frequencies are known, the quantity which can be estimated with this
setup is therefore given by $\Omega\equiv(\omega_1+\omega_2)/2$ and
the corresponding precision, which can be calculated via the QFI, is
given by $\Delta \Omega_{min} = 1/\sqrt{Tt}N$ which has Heisenberg
scaling. The above discussion can again be generalized to an arbitrary
ordering of the atoms as long as we use a GHZ state as input.  The
setup can be realized, e.g., by using transitions
$\ket{m}\leftrightarrow\ket{\tilde m}$ with frequency $\omega_1$ and
$\ket{-m}\leftrightarrow\ket{-\tilde m}$ with frequency $\omega_2$,
i.e. the two ground states (with magnetic quantum numbers $\pm m$) and
the two excited states (with magnetic quantum numbers $\pm \tilde m$)
are in the same Zeeman manifold, respectively, such that (fluctuating)
magnetic fields cause first order Zeeman shifts of the same magnitude
but opposite sign~\cite{Roos05,Chwalla07}.  Note that the quantity to be
estimated, $\Omega$, is magnetic field independent (in first order),
i.e. the situation is similar to a clock transition, and $\Omega$
might therefore serve as a frequency standard. Moreover, $\Omega$ can
easily be chosen to be in the optical domain which is desirable for
atomic clocks~\cite{Diddams04}.  

We also note that, similar to the case of estimating $\delta$, we can
introduce decoherence free logical states $\ket{0}_L\equiv\ket{00}$
and $\ket{1}_L\equiv\ket{11}$ which accumulate a relative phase
$2\Omega t$. A GHZ state of $N$ atoms is then simply a GHZ state of
$n=N/2$ logical states and the precision is given by
$\Delta\Omega_{min}=1/\sqrt{Tt}2n$, the factor of 2 arising from the
factor of 2 in the relative phase of $\ket{0}_L$ and $\ket{1}_L$.

\begin{figure}[t!]
\centering\includegraphics[]{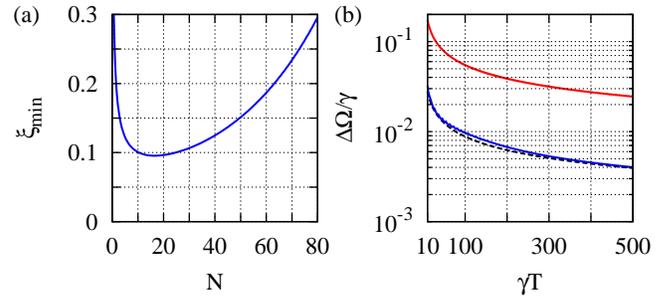}
\caption{%
  (Color online) (a) Minimum fidelity of the input state versus number
  of atoms $N$ [cf. Eq.~(\ref{eq:fid_bound})]. (b) Maximum likelihood
  estimation uncertainty $\Delta\Omega$ versus total measurement time
  $T$ for $N=20$ and $\xi(N)=0.6$. The lower, solid (blue) line
  corresponds to a GHZ state and the upper, solid (red) line to a
  product state. The dashed (black) line is given by
  Eq.~(\ref{eq:prec}).  In both figures we set $\eta_H=0.98$,
  $\eta_M=0.99$, $\gamma t=3$.}
\label{fig3}
\end{figure}

An optimal measurement for the two schemes discussed above, i.e. a
measurement for which $F=F_Q$ [see Eq.~(\ref{eq:crb})], is given by 
a $\pi/2$-pulse and a measurement of each atom in the
$\{\ket{0},\ket{1}\}$-basis [see Fig.~\ref{fig1}(b)]. However, in
practice there will be imperfections both in the preparation of the
input state and the measurement.  A faulty measurement of an atom can
be modeled using the measurement operators 
\begin{equation}
\Pi_i =\frac{1}{2}(1+\eta_M)\ket{i}\bra{i} +
\frac{1}{2}(1-\eta_M)\sigma_x\ket{i}\bra{i}\sigma_x,
\end{equation}
where $i=0,1$,
and $\eta_M$ is the likeliness that we get the correct measurement
result.  Also, we assume that the actual input state is of the form
\begin{equation}
\rho_{in} = \xi(N) \ket{\psi_{in}}\bra{\psi_{in}} + \frac{1}{2^N}[1-\xi(N)]\openone,
\label{eq:err} 
\end{equation}
i.e. we prepare the ideal input state $\ket{\psi_{in}}$ with fidelity
\begin{equation}
f = \bra{\psi_{in}}\rho_{in}\ket{\psi_{in}} = \xi(N)+\frac{1}{2^N}[1-\xi(N)], 
\end{equation}
and hence $f\approx \xi(N)$ for $N\gg 1$.  The error
model~(\ref{eq:err}) is a worst case scenario since the identity
matrix does not yield any phase information. Analogously, we assume
that a $\pi/2$-pulse is given by the operation
\begin{equation}
  \eta_H \rho_{id} + \frac{1}{2}(1-\eta_H)\openone,
\end{equation}
i.e. $\eta_H$ characterizes the probability to perform a perfect
$\pi/2$-pulse leading to an ideal state $\rho_{id}$ (see Appendix~\ref{sec:AppIII}).  
We can then calculate the Fisher information for both schemes and
therefore the Cram\'er-Rao bounds,
\begin{equation}
\Delta\Omega_{min} = \frac{\Delta\delta_{min}}{2} = \frac{1}{\sqrt{Tt}N\xi(N)\eta_H^N\eta_M^N},
\label{eq:prec}
\end{equation} 
where we assumed that $\varphi_1\pm\varphi_2=\pi/N$ (`+' for
$\Delta\Omega_{min}$; `-' for $\Delta\delta_{min}$), which can always
be achieved by a feedback setup which appropriately adjusts, e.g., the
electric quadrupole field, laser frequencies or/and the evolution time
$t$. We note that the state which was prepared in~\cite{Chwalla07}
leads to the same result for $\xi(N)=1/2$ and $N=2$.  The term
$\xi(N)\eta_H^N\eta_M^N$ in Eq.~(\ref{eq:prec}) might, at first
glance, lead to the conclusion that faulty state preparation and
detection annihilates the advantage gained by using a DFS.  To show
that this is not the case we compare the precisions~(\ref{eq:prec}) to
those obtained using conventional Ramsey spectroscopy with a
product-state $\ket{\psi_{in}^{pro}}$ as input. In this case we would
use $N/2$ atoms to estimate $\omega_1$ and the others to estimate
$\omega_2$.  For a fair comparison we assume uncorrelated dephasing
which can in principle always be achieved by placing the atoms in
different traps.  Assuming that $\ket{\psi_{in}^{pro}}$ is created by
$N$ $\pi/2$-pulses, the corresponding precision then reads
$\Delta\omega_{1,min}=\Delta\omega_{2,min}=\sqrt{4\e\gamma/NT}/\eta_H^2\eta_M$,
and the precisions of the quantities to be estimated is given by
$\Delta\Omega_{min}=\frac{1}{2}\Delta\delta_{min} \approx
\Delta\omega_{1,min}/\sqrt2$ which have to be compared to
Eq.~(\ref{eq:prec}). In both cases this leads to the constraint
\begin{equation}
\xi(N) \ge \xi_{min} \equiv \frac{1}{\eta_H^{N-2}\eta_M^{N-1}\sqrt{2N\gamma t\e}},
\label{eq:fid_bound}
\end{equation}
i.e. whenever the above inequality is fulfilled the DFS schemes beat
conventional Ramsey spectroscopy.  An example is shown in
Fig.~\ref{fig3}(a).  With current ion trap experiments gate and
readout fidelities in excess of $\eta_H=0.98$ and $\eta_M=0.99$ have
been achieved~\cite{Lucas10}.  Furthermore, we assumed that using a
DFS scheme leads to a coherence time which is 3 times longer than the
coherence time of a single atom. This is a rather conservative
estimate which has already been exceeded in experiments~\cite{Monz10}.
As can be seen the bound for the state fidelity $\xi(N)$ is
surprisingly low. In the experiment described in~\cite{Monz10} a
50.8\% fidelity for a $N=14$ GHZ state was achieved. For our scheme it
would be required to manipulate this GHZ state by transfering half of
the atoms into a different internal state. Naturally this would be
done by addressing, for example, the second $N/2$ neighbouring atoms
by an appropriate sequence of laser pulses, i.e., crucially, it is not
required to address atoms individually. This would of course decrease
$\xi(N)$ but even if it reduces it to, say, 20\% (which is a very
conservative assumption) we still beat conventional Ramsey
spectroscopy.

The bounds~(\ref{eq:prec}) can be reached using maximum likelihood estimation in the
limit of large $\nu$ (or $T$). In practice it is certainly highly
relevant {\em how} large $\nu$ has to be such that the actual
estimation uncertainty is close to the bound. Suppose we perform $\nu$
experimental runs and obtain the results $n_1,\ldots, n_\nu$, where
$n_j$ is the number of times the state $\ket{0}$ is measured in each
run, and the total number of even $n_j$ is $\nu_e$.  It turns out that
the maximum likelihood estimators $\Omega_{est}$ and $\delta_{est}$ depend only on
$\nu_e$ (see Appendix~\ref{sec:AppIV}). Using the probability distribution for
$\nu_e$ and Eq.~(\ref{eq:uncertainty}) we can then calculate, e.g. the
estimation uncertainty $\Delta\Omega$ for finite $\nu$.  A result is
shown in Fig.~\ref{fig3}(b) depending on the total time $T=\nu t$ of
the experiment (lower solid line) for $\varphi_1+\varphi_2=\pi/N$.  As
can be seen the estimation uncertainty quickly approaches the lower
bound (dashed line). We also show the estimation uncertainty and the
lower bound for conventional Ramsey spectroscopy with
$\ket{\psi_{in}^{pro}}$ (upper solid line; the two quantities are
indistinguishable on the scale of the figure).  Evidently, even for
small $T$, the DFS scheme easily outperforms conventional Ramsey
spectroscopy. We note that the corresponding plots for estimating
$\delta$ would be identical to the ones shown but larger by a factor
of two.

To conclude, we have shown that correlated dephasing significantly
diminishes the precision of frequency estimation with standard Ramsey
interferometry. On the other hand, it allows for the existence of DFSs
which we used to construct and analyze generalized Ramsey setups which
beat the SQL even in the presence of faulty detection and
significantly imperfect state preparation. The proposed schemes for
quantum enhanced frequency estimation are therefore feasible with
current experimental technology and can lead to improved spectroscopic
methods with a variety of important applications in metrology.

\acknowledgments
We acknowledge support for this work by the National Research
Foundation and Ministry of Education, Singapore and Keble College, Oxford.

\appendix

\section{Noise model}
\label{sec:AppI}
In this appendix we give a detailed description of the noise model and
the derivation of Eq.~(\ref{eq:dyn1}). An alternative derivation is given by
coupling the atoms to a bosonic bath, similar to the methods described
in~\cite{Palma96,Carmichael}.  However, the following derivation,
which is based on a Langevin-equation approach, is physically more
intuitive for the systems considered in this paper, i.e. atoms which
are subject to fluctuating classical fields.

Consider $N$ two-level atoms (or ions) with internal states
$\{\ket{0}_j,\,\ket{1}_j\}$ and transition frequencies $\omega_j$
which are subject to a time-dependent, fluctuating field leading to
random energy shifts of the transitions. The Hamiltonian can then be
written as
\begin{eqnarray}
H &=& \frac{1}{2}\sum_{j=1}^N \omega_j \sigma_z^j + B(t) \sum_{j=1}^N \varepsilon_j \sigma_z^j \nonumber\\
&\equiv& H_0 + B(t)L,
\end{eqnarray}
where (the real numbers) $B(t)$ and $\varepsilon_j$ characterize the
field strength and how it affects atom $j$, and $\sigma_z^j=\ket{0}_j\bra{0}-\ket{1}_j\bra{1}$ is the
Pauli $z$-operator acting on atom $j$. Writing $B(t)\equiv
\sqrt{\gamma/2}\xi(t)$ the Schr\"odinger equation then takes the form
\begin{equation}
\frac{d}{dt}\ket{\psi} = -\ii H_0\ket{\psi} - \ii\sqrt{\frac{\gamma}{2}} L \ket{\psi}\xi(t).
\label{eq:lang}
\end{equation}
The random fluctuations of the field are captured in $\xi(t)$ and
Eq.~({\ref{eq:lang}}) is therefore an example of a Langevin
equation~\cite{Gardiner}. Assuming that $\xi(t)$ has zero mean
and very rapidly decaying time correlations, we can write
\begin{eqnarray}
\overline{\xi(t)} &=& 0,\\
\overline{\xi(t)\xi(t')} &=& \delta(t-t') \label{eq:corr},
\end{eqnarray}
where the overbar denotes the mean value, i.e. we make the idealization
that $\xi(t)$ is white noise. Equation~(\ref{eq:lang}) can then be written as
a stochastic differential equation
\begin{equation}
\ket{d\psi} = -\ii H_0 \ket{\psi}dt -\ii  \sqrt{\frac{\gamma}{2}} L \ket{\psi}dW\qquad(S),
\label{eq:strat}
\end{equation}
where $dW = \xi(t)dt$ is a Wiener increment~\cite{Gardiner}. The $(S)$
indicates that we have to interpret this equation in the Stratonovich
sense. The reason for this is given by the fact that Eq.~(\ref{eq:corr})
is an idealization. In reality this correlation function will have
a finite width (but which is small compared to any other relevant time
scale). In such a case the stochastic differential equation
(\ref{eq:strat}) has to be interpreted in the Stratonovich sense. A
detailed discussion of this point can be found in Chapter 6.5
in~\cite{Gardiner}. Transforming Eq.~(\ref{eq:strat}) into Ito form, using the standard
rules~\cite{Gardiner}, leads to
\begin{equation}
\ket{d\psi} =  \left( -\ii H_0  - \frac{\gamma}{4} L^2 \right)\ket{\psi}dt -\ii\sqrt{\frac{\gamma}{2}}L \ket{\psi}dW\qquad(I),
\label{eq:ito}
\end{equation}
where the $(I)$ indicates that this equation is to be interpreted 
in the Ito sense. Using Ito calculus, the time evolution for the
density operator $\varrho\equiv\ket{\psi}\bra{\psi}$ is then derived
to be
\begin{eqnarray}
d\varrho &=& \varrho(t+dt) - \varrho(t) \nonumber\\
&=& \ket{\psi(t+dt)}\bra{\psi(t+dt)} - \ket{\psi(t)}\bra{\psi(t)} \nonumber\\
&=& (\ket{\psi(t)} + \ket{d\psi})(\bra{\psi(t)} + \bra{d\psi}) - \ket{\psi(t)}\bra{\psi(t)} \nonumber\\
&=& \ket{d\psi}\bra{\psi} + \ket{\psi}\bra{d\psi} + \ket{d\psi}\bra{d\psi} \nonumber\\
&=& \left( -\ii H_0  - \frac{\gamma}{4} L^2 \right)\varrho dt 
+\varrho\left( \ii H_0  - \frac{\gamma}{4} L^2 \right) dt  \nonumber\\
&& -\ii\sqrt{\frac{\gamma}{2}}L \varrho dW +\ii\sqrt{\frac{\gamma}{2}} \varrho L dW + \frac{\gamma}{2} L\varrho L dt \qquad (I). \nonumber\\
\end{eqnarray}
For the averaged density operator
\begin{equation}
\rho(t) \equiv \overline{ \varrho(t) }
\end{equation}
we therefore get
\begin{equation}
\dot\rho = -\ii[H_0,\rho] + \frac{\gamma}{2}\left(  L\rho L - \frac{1}{2}L^2\rho - \frac{1}{2}\rho L^2   \right).
\label{eq:master}
\end{equation}
In this paper we consider three different scenarios
corresponding to special cases of the above equation, some of them
make use of decoherence free subspaces (DFSs):

\subsection{Conventional Ramsey Spectroscopy}
\label{sec:conv}
In conventional Ramsey spectroscopy we consider the same transition in
each atom, i.e.  $\omega=\omega_j,\,j=1\ldots N$ and hence all
transitions will be affected in the same way by the fluctuating field,
i.e. $\varepsilon_j=1,\,j=1\ldots N$. It follows that we have (in a
rotating frame with respect to a laser frequency $\omega_L$)
\begin{equation}
L = S_z \equiv \sum_{j=1}^N \sigma_z^j,\quad
H_0 = \frac{(\omega-\omega_L)}{2} S_z,
\label{eq:conv}
\end{equation}
and thus Eq.~(\ref{eq:master}) is equal to Eq.~(\ref{eq:dyn1}).

\subsection{DFS spectroscopy for estimating difference of two frequencies.} 
\label{sec:DFS1}
Here we assume that half of the atoms, represented by a set $A$, have
transition frequency $\omega_1$ and the other half, represented by a
set $B$, have transition frequency $\omega_2$ and the fluctuating
field leads to the same energy shift in both transitions, i.e.
$\varepsilon_j=1,\,j=1\ldots N$. Hence we obtain
\begin{eqnarray}
L &=& S_z,\nonumber\\
H_0 &=& \frac{\omega_1}{2} \sum_{j\in A} \sigma_z^j +\frac{\omega_2}{2}\sum_{j\in B} \sigma_z^j \nonumber\\
&=& \frac{\omega_1-\omega_2}{4}\left( \sum_{j\in A} \sigma_z^j - \sum_{j\in B} \sigma_z^j  \right)
+\frac{\omega_1+\omega_2}{4} L. \qquad
\end{eqnarray}
If we use a state of the form
\begin{equation}
\ket{\psi_{in}} = \frac{1}{\sqrt 2}( \ket{i_1,i_2,\ldots,i_N} +
\prod_{j=1}^N\sigma_x^j\ket{i_1,i_2,\ldots,i_N} ),
\label{eq:state1}
\end{equation}
where  $i_j=0$ if $j\in A$, $i_j=1$ if $j\in B$ and $\sigma_x^j=\ket{0}_j\bra{1}+\ket{1}_j\bra{0}$, we have
$L\ket{\psi_{in}}=0$ and the above setup can be used for the estimation of
$\delta=\omega_1-\omega_2$.

\subsection{DFS spectroscopy for estimating mean of two frequencies.} 
\label{sec:DFS2}
Finally, we assume that half of the atoms, represented by a set $A$,
have transition frequency $\omega_1$ and the other half, represented
by a set $B$, have transition frequency $\omega_2$ and the fluctuating
field leads to energy shifts of the two transitions of the same
magnitude but opposite sign.  More exactly,
$\omega_j=\omega_1,\,\varepsilon_j=-1$ if $j\in A$ and
$\omega_j=\omega_2,\,\varepsilon_j=1$ if $j\in B$.  In this case we
obtain (in a rotating frame with respect to laser frequencies
$\omega_{L1}$ and $\omega_{L2}$)
\begin{eqnarray}
L &=& -\sum_{j\in A} \sigma_z^j + \sum_{j\in B} \sigma_z^j,\\
H_0 &=& \frac{(\omega_1-\omega_{L1})}{2} \sum_{j\in A} \sigma_z^j + \frac{(\omega_2-\omega_{L2})}{2} \sum_{j\in B} \sigma_z^j \nonumber\\
&=& \frac{\delta_1+\delta_2}{4}S_z + \frac{\delta_2-\delta_1}{4}L,
\end{eqnarray}
where $\delta_i=\omega_i-\omega_{Li}$. An $N$-particle GHZ state
\begin{equation}
\ket{\psi_{in}} = \frac{1}{\sqrt 2}( \ket{00\ldots0} + \ket{11\ldots1} )
\label{eq:state2}
\end{equation}
has the property $L\ket{\psi_{in}}=0$. This setup can be used to
estimate $\Omega=(\omega_1+\omega_2)/2$.

\section{Quantum Fisher information}
\label{sec:AppII}
Consider a system state $\rho$ which depends on a parameter $\alpha$
which is to be estimated. Defining $\rho'\equiv\frac{d}{d\alpha}\rho$,
the quantum Fisher information (QFI) is given by
\begin{equation}
F_Q=Tr\{\rho'\mathcal{L}_\rho(\rho')\},
\end{equation}
where $\mathcal{L}_\rho(\rho')$ is the ``symmetric logarithmic
derivative'' (SLD) of
$\rho$~\cite{Helstrom,Holevo,Braunstein94,Braunstein96}. Writing the
state in diagonal form, $\rho=\sum_{j}p_j\ket{\psi_j}\bra{\psi_j}$,
the SLD is given by
\begin{equation}
\mathcal{L}_\rho(\rho') = \sum_{j,k;\,p_k+p_j\ne0}\frac{2}{p_j+p_k}\bra{\psi_j}\rho'\ket{\psi_k}\ket{\psi_j}\bra{\psi_k}
\end{equation}
and therefore
\begin{equation}
F_Q = \sum_{j,k;\,p_k+p_j\ne0}\frac{2}{p_j+p_k}|\bra{\psi_j}\rho'\ket{\psi_k}|^2.
\end{equation}
Assume that $\rho$ is the solution of Eq.~(\ref{eq:master}) and
$[L,H_0]=0$ and that only $H_0$ depends on the parameter $\alpha$ such
that $[H_0',H_0]=0$, where $H_0'\equiv\frac{d}{d\alpha}H_0$. We then
obtain $\rho=\e^{-\ii H_0t}\tilde\rho\e^{\ii H_0t}$, where
$\tilde\rho$ is the solution of Eq.~(\ref{eq:master}) with
$H_0\equiv0$, and therefore
\begin{equation}
F_Q = 2t^2\sum_{j,k}\frac{(p_j-p_k)^2}{p_j+p_k}|\bra{\psi_j}H_0'\ket{\psi_k}|^2.
\label{eq:QFI}
\end{equation}
Note that if $\rho=\ket{\psi}\bra{\psi}$ is a pure state, the above
reduces to $F_Q=4t^2(\bra{\psi} (H_0')^2\ket{\psi} - \bra{\psi}
H_0'\ket{\psi}^2 )$.

In conventional Ramsey spectroscopy (see Appendix~\ref{sec:conv}) a
GHZ state of the form~(\ref{eq:state2}) would evolve into
\begin{align}
  &\rho(t)=\frac{1}{2}\Big[
  \ket{0\ldots0}\bra{0\ldots0} +\ket{1\ldots1}\bra{1\ldots1}      \nonumber\\
  &+ \e^{-\gamma N^2t}\big(\e^{-\ii\delta Nt}
  \ket{0\ldots0}\bra{1\ldots1} + \e^{\ii\delta
    Nt}\ket{1\ldots1}\bra{0\ldots0} \big) \Big],
\label{eq:ghz_deco}
\end{align}
where $\delta=\omega-\omega_L$. As can be seen, due to correlated
dephasing, the above state decoheres on a timescale $1/\gamma N^2$
which is shorter than the decoherence timescale in the presence of
uncorrelated dephasing (given by $1/\gamma N$). This behavior was
therefore dubbed ``superdecoherence''~\cite{Palma96,Monz10}. The
parameter to be estimated is the transition frequency $\alpha=\omega$.
The corresponding QFI is obtained by diagonalizing the
state~(\ref{eq:ghz_deco}) and using Eq.~(\ref{eq:QFI}) leading to
$F_Q=t^2N^2\e^{-2\gamma N^2t}$. The corresponding precision
\begin{equation}
\Delta\omega_{min} = \frac{1}{\sqrt{\nu F_Q}}= \frac{1}{\sqrt{Tt}N\e^{-\gamma N^2t}}
\end{equation}
is optimal for a time $t_{opt}=1/2\gamma N^2$ leading to
$\Delta\omega_{min}^{opt}=\sqrt{2\e\gamma/T}$ which, as a consequence
of superdecoherence, has no $N$ dependency.  GHZ states are therefore
not particularly useful for conventional Ramsey spectroscopy.

To find the best possible precision in conventional Ramsey interferometry we can
restrict ourselves to input states which are symmetric under exchange of
particles. To prove this we use a method inspired by
Ref.~\cite{Buzek99}. Consider a unitary operation $U$ which maps an
arbitrary state $\ket{\psi}$ onto a symmetric state, i.e.
\begin{equation}
U\ket{\psi} = \sum_{k=0}^N c_k \ket{k,N-k}.
\label{eq:U}
\end{equation}
The state $\ket{k,N-k}$ is the completely symmetrized state with $k$
atoms in state $\ket{0}$ and $N-k$ atoms in state $\ket{1}$ and is
defined by
\begin{equation}
\ket{k,N-k} = \sqrt{\binom{N}{k}^{-1}}\sum_P P \ket{i_1,i_2,\ldots,i_N},
\label{eq:fockstate}
\end{equation}
where $i_j=0,1$ and $k$ ($N-k$) is the number of zeros (ones) in
$\ket{i_1,i_2,\ldots,i_N}$. Furthermore, the sum is over all permutations $P$ of
particles which lead to different terms in the sum. Before we proceed
we will show $(i)$ that such an $U$ always exists and $(ii)$ that
$[U,S_z]=0$.

{\em Proof of (i)}: An arbitrary state of the system can be written as
\begin{equation}
\ket{\psi} = \sum_{k=0}^N \sum_{\mu=1}^{M_k} b_{k,\mu} \ket{k;\mu},
\end{equation}
where $\ket{k;\mu} = \ket{i_1,i_2,\ldots,i_N}$, $i_j=0,1$ and $k$ is
the number of times $i_j$ is zero. For each $k$ there are
$M_k=\binom{N}{k}$ such states which we enumerate using $\mu$. On the
subspace defined by a fixed $k$ we define a basis
$\{\ket{\phi_j(k)}\,|\,j=1,\ldots,M_k\}$ such that
$\ket{\phi_1(k)}\equiv \frac{1}{\mathcal{N}_k}\sum_{\mu=1}^{M_k}
b_{k,\mu} \ket{k;\mu}$ with $\mathcal{N}_k=\sqrt{\sum_{\mu=1}^{M_k}
  |b_{k,\mu}|^2}$ and the remaining $\{\ket{\phi_j(k)}\,|\,j>1\}$ are
chosen such that we obtain an orthonormal basis. Equation~(\ref{eq:U})
is fulfilled if
$U\ket{\phi_1(k)}=\frac{c_k}{\mathcal{N}_k}\ket{k,N-k}$ for all $k$.
Decomposing $U$ into a block diagonal form $U=\oplus_{k=0}^N \,U(k)$,
i.e.  $U(k)$ is the $M_k\times M_k$ block acting on the subspace $k$,
the above can be achieved by requiring that the first column of the
matrix $U(k)$ has elements $U_{j1}(k)=\bra{\phi_j(k)}U(k)
\ket{\phi_1(k)} = \frac{c_k}{\mathcal{N}_k}\langle
\phi_j(k)|k,N-k\rangle$, $j=1,\ldots,M_k$.  The remaining columns can
be chosen such that all columns are mutually orthonormal. If
$\mathcal{N}_k$ is zero for some $k$ we can choose $U(k)$ to be the
identity. Therefore $U(k)$ and hence $U$ can be chosen to be unitary.

{\em Proof of (ii)}: We have 
\begin{align}
S_zU\ket{\psi} &= S_z\sum_{k=0}^N c_k \ket{k,N-k} \nonumber \\
&= \sum_{k=0}^N (2k-N)c_k\ket{k,N-k} \nonumber\\
&= \sum_{k=0}^N (2k-N)  U\,\sum_{\mu=1}^{M_k} b_{k,\mu} \ket{k;\mu} = US_z\ket{\psi}
\end{align}
and hence $[U,S_z]=0$.

We can now use $(i)$ and $(ii)$ to show that to find the optimal
precision we can restrict ourselves to the symmetric subspace.
Consider an arbitrary pure input state $\ket{\psi_{in}}$. The time
evolution of this state in conventional Ramsey interferometry as
defined in Appendix~\ref{sec:conv} and Eq.~(\ref{eq:dyn1}) is given by
\begin{align}
\rho(t) = 
&\e^{-\ii \frac{\delta}{2} S_z t}
\e^{-\frac{\gamma}{4}S_z^2t} \nonumber\\
&\times\sum_{m=0}^\infty \frac{(\gamma t/2)^m}{m!} S_z^m \ket{\psi_{in}}\bra{\psi_{in}} S_z^m 
\e^{-\frac{\gamma}{4}S_z^2t}
\e^{\ii \frac{\delta}{2} S_z t}.
\end{align}
If we take the symmetric state $U\ket{\psi_{in}}$ as input state
instead of $\ket{\psi_{in}}$ the state of the system at time $t$ has
the form $\rho^s(t)=U\rho(t)U^\dagger$ due to $(ii)$. Diagonalizing
$\rho(t)$ using an orthonormal basis we can write $\rho(t) =
\sum_{k}p_k\ket{\psi_k}\bra{\psi_k}$ and therefore $\rho^s(t) =
\sum_{k}p_k\ket{\psi_k^s}\bra{\psi_k^s}$ with
$\ket{\psi_k^s}=U\ket{\psi_k}$. Due to $(ii)$ the QFIs of $\rho^s(t)$
and $\rho(t)$ are therefore equal,
\begin{align}
  F_Q[\rho^s] &= 2t^2\sum_{j,k}\frac{(p_j-p_k)^2}{p_j+p_k}|\bra{\psi_j^s}S_z\ket{\psi_k^s}|^2 \nonumber \\
  &=
  2t^2\sum_{j,k}\frac{(p_j-p_k)^2}{p_j+p_k}|\bra{\psi_j}S_z\ket{\psi_k}|^2
  = F_Q[\rho].
\label{eq:QFI_equal}
\end{align}
Assuming that $\ket{\psi_{in}}$ is an optimal input state which
maximizes the QFI then $U\ket{\psi_{in}}$ is optimal as well and
therefore we can restrict our search to the symmetric subspace which
concludes the proof.

The Fock states defined by Eq.~(\ref{eq:fockstate}) represent states
with $k$ atoms in state $\ket{0}$ and $N-k$ atoms in state
$\ket{1}$. Also the operator $S_z$ can be written in a Fock
representation given by $S_z=n_0-n_1$, where $n_i=a_i^\dagger a_i$ and
$a_i$ ($a_i^\dagger$) are bosonic annihilation (creation) operators
for modes $i=0,1$. Since the total particle number $n_0+n_1=N$ is
conserved we can set $S_z=2n_0-N$ and therefore Eq.~(\ref{eq:dyn1}) transforms into
\begin{equation}
\dot\rho = -\ii\delta[n_0,\rho] + 2\gamma\left(  n_0\rho n_0 - \frac{1}{2}n_0^2\rho - \frac{1}{2}\rho n_0^2   \right).
\label{eq:master2}
\end{equation}
Furthermore, every symmetric, pure input state can be written in the form
\begin{equation}
\ket{\psi_{in}} = \sum_{k=0}^N \alpha_k \ket{k,N-k}.
\end{equation}
The solution of Eq.~(\ref{eq:master2}) is given by
\begin{align}
\rho(t) &= \e^{-\ii\delta n_0t} \e^{-\gamma n_0^2 t}\sum_{m=0}^\infty\frac{(2\gamma t)^m}{m!}n_0^m \rho(0) n_0^m \e^{-\gamma n_0^2 t}\e^{\ii\delta n_0t} \nonumber\\
 &= \sum_{k,l=0}^N \alpha_k\alpha_l^* \e^{-\gamma
  t(k-l)^2}\e^{-\ii\delta t(k-l)} \ket{k,N-k}\bra{l,N-l},
\label{eq:rho}
\end{align}
where we set $\rho(0)=\ket{\psi_{in}}\bra{\psi_{in}}$. In order to
find the input state which leads to the best possible precision for
estimating $\omega$ we performed a numerical optimization in the
bosonic picture, where the time evolution is given by
Eq.~(\ref{eq:master2}) using methods described in~\cite{Press} and a
result is shown in Fig.~\ref{fig2}(b).

A product state
\begin{equation}
  \ket{\psi_{in}^{pro}}= \left[\frac{1}{\sqrt2}(\ket{0}+\ket{1})\right]^{\otimes N}
  \label{eq:pro}
\end{equation}
is symmetric under particle exchange, and in the Fock representation
it takes the form of a `coherent state'
\begin{eqnarray}
  \ket{\psi_{in}^{pro}} &=& \frac{1}{2^\frac{N}{2}}\sum_{k=0}^N \sqrt{\binom{N}{k}} \ket{k,N-k} \nonumber\\
  &=& \frac{1}{\sqrt{2^N N!}}(a_0^\dagger + a_1^\dagger)^N\ket{0,0}.
\end{eqnarray}
Using this as input state, the density matrix~(\ref{eq:rho}) can be
numerically diagonalized, and via Eq.~(\ref{eq:QFI}) we calculate the
QFI. Like for a GHZ state, $\Delta\omega_{min}$ can be minimized for a
time $t_{opt}$ and the corresponding precision is obtained to be
$\Delta\omega_{min}^{opt} \approx (\sqrt{2} +
0.87/N^{0.90})\sqrt{\gamma/T}$. This shows that, also in the case of
product states, correlated dephasing is more detrimental than
uncorrelated dephasing (in which case we would obtain
$\Delta\omega_{min}^{opt} =
\sqrt{2\e/N}\sqrt{\gamma/T}$~\cite{Huelga97}). Furthermore, the best
possible precision is only marginally better than the precision
obtained by using a product state [see Fig.~\ref{fig2}(b)] showing
that conventional Ramsey spectroscopy is merely of limited use for
frequency estimation in the presence of collective dephasing.

\section{Fisher information}
\label{sec:AppIII}
The QFI provides the optimal precision for estimating a parameter. It
depends only of the system state before the measurement and not on the
measurement itself. In order to examine the effects of particular measurements on the estimation precision we
therefore have to consider the Fisher information (FI). The FI is
given by
\begin{equation}
F = \sum_k \frac{1}{p(k|\alpha)}\left(\frac{d}{d\alpha}p(k|\alpha)\right)^2,
\end{equation}
where $p(k|\alpha)$ is the probability to obtain a measurement outcome
$k$ given that the value of the parameter to be estimated is $\alpha$,
\begin{equation}
p(k|\alpha) = Tr\{ \Pi_k \rho(\alpha)  \}.
\end{equation}
Here, the operators $\Pi_k$ form a positive operator valued measure
(POVM) describing the measurement. If for a particular POVM the FI is
equal to the QFI the measurement is said to be optimal, i.e. it
saturates the quantum Cram\'er-Rao bound [see Eq.~(\ref{eq:crb})].

Both for the estimation of $\alpha=\delta=\omega_1-\omega_2$ and
$\alpha=\Omega=(\omega_1+\omega_2)/2$, i.e. the two schemes described
in Appendices~\ref{sec:DFS1} and~\ref{sec:DFS2}, the optimal
measurement is given by a measurement of all atoms in the
$\sigma_x$-basis which in practice is done by a Hadamard gate and a
measurement in the $\{\ket{0},\,\ket{1}\}$-basis. Note that we use
Hadamard gates for simplicity. In practice these can be replaced by
$\pi/2$-pulses which has no effect on the FI. In an actual experiment
both Hadamard gate and measurement will have imperfections. To model
these we assume that the Hadamard operation on one atom is given by
\begin{equation}
  \mathcal{E}_H(\rho) = \eta_H H\rho H + \frac{1}{2}(1-\eta_H)\openone,
\end{equation}
where $H$ is a perfect Hadamard gate and $\eta_H$ characterizes the
probability to have a perfect gate. It corresponds to the gate fidelity
$f_H$, as defined e.g. in~\cite{Nielsen}, via
$f_H=\sqrt{(1+\eta_H)/2}$. The POVM for the measurement of one atom is
given by
\begin{eqnarray}
  \Pi_0 &=& \frac{1+\eta_M}{2} \ket{0}\bra{0} + \frac{1-\eta_M}{2} \ket{1}\bra{1},\nonumber\\
  \Pi_1 &=& \frac{1+\eta_M}{2} \ket{1}\bra{1} + \frac{1-\eta_M}{2} \ket{0}\bra{0},
\end{eqnarray}
i.e. $\eta_M$ quantifies the probability that we have a perfect
measurement. The above can be combined into a new POVM which describes
a faulty measurement in the $\sigma_x$-basis,
\begin{equation}
  \Pi_\pm = \frac{1+\eta_H\eta_M}{2} \ket{\pm}\bra{\pm} + \frac{1-\eta_H\eta_M}{2} \ket{\mp}\bra{\mp},
  \label{eq:POVMX}
\end{equation}
where $\ket{\pm}=(\ket{0}\pm\ket{1})/\sqrt2$ are eigenstates of
$\sigma_x$. Imperfect state preparation can be modeled by
\begin{equation}
  \rho_{in} = \xi(N)\ket{\psi_{in}}\bra{\psi_{in}} + \frac{1}{2^N}[1-\xi(N)]\openone,
\end{equation}
where $\ket{\psi_{in}}$ is the ideal, pure input state. For the
estimation of $\alpha=\delta=\omega_1-\omega_2$ (see
Appendix~\ref{sec:DFS1}) the state $\ket{\psi_{in}}$ is given by
Eq.~(\ref{eq:state1}) and for the estimation of
$\alpha=\Omega=(\omega_1+\omega_2)/2$ (see Appendix~\ref{sec:DFS2})
the state $\ket{\psi_{in}}$ is given by Eq.~(\ref{eq:state2}). The
state $\rho_{in}$ evolves then into the state $\rho(\alpha)$, the
state before the measurement, according to the dynamics given by the
Hamiltonians in Appendix~\ref{sec:DFS1} and Appendix~\ref{sec:DFS2},
respectively.

A particular outcome $k$ of a measurement on all $N$ atoms is given by
a sequence $\{i_1,i_2,i_3,\ldots,i_N\}$ where $i_j=\pm$, i.e. if the
$j$th atom is found in state $\ket{+}$ ($\ket{-}$) we have $i_j=+$
($i_j=-$).  Note that in practice a $\ket{+}$ ($\ket{-}$) outcome
corresponds to finding the atoms in state $\ket{0}$ ($\ket{1}$) due to
the Hadamard gate. The probability for a particular outcome is then
calculated to be
\begin{eqnarray}
p(k|\alpha) &=& Tr\{ \Pi_{i_1}\Pi_{i_2}\Pi_{i_3}\ldots\Pi_{i_N} \rho(\alpha) \}\nonumber\\
&=& \frac{1}{2^N}\left( 1+(-1)^{n}  \xi(N)\eta_H^N\eta_M^N\cos[N\varphi(\alpha)]\right) \nonumber\\
&\equiv& q_n(\alpha),
\label{eq:prob}
\end{eqnarray}
where
\begin{equation}
\varphi(\alpha)=
\begin{cases}
  [\Omega-(\omega_{L1}+\omega_{L2})/2]t &; \text{ if }\alpha=\Omega \\
  \delta t/2 &; \text{ if } \alpha=\delta,
\end{cases}
\label{eq:phialphas}
\end{equation}
and $n$ is the number of times `+' is contained in
the sequence $\{i_1,i_2,i_3,\ldots,i_N\}$. 
From this we obtain the FI
\begin{eqnarray}
F &=& \sum_{n=0}^N \binom{N}{n}\frac{1}{q_n(\alpha)}\left( \frac{d}{d\alpha}q_n(\alpha)\right)^2 \nonumber\\
&=& \frac{\left(c(\alpha)Nt\xi(N)\eta_H^N\eta_M^N\right)^2\sin^2[N\varphi(\alpha)]}{1-\left(\xi(N)\eta_H^N\eta_M^N\right)^2\cos^2[N\varphi(\alpha)]},\nonumber \\
\end{eqnarray}
where $c(\alpha=\delta)=1/2$ and $c(\alpha=\Omega)=1$. The FI is
maximized for $\varphi(\alpha) = \pi/2N$ leading to the Cram\'er-Rao bound
\begin{equation}
\Delta\Omega_{min} = \frac{\Delta\delta_{min}}{2}=\frac{1}{\sqrt{\nu}tN\xi(N)\eta_H^N\eta_M^N}. 
\label{eq:FI}
\end{equation}
Setting $\nu=T/t$ we therefore obtain Eq.~(\ref{eq:prec}).

Equation~(\ref{eq:FI}) has to be compared to the precision
corresponding to `classical' Ramsey spectroscopy using the product
state~(\ref{eq:pro}) as input, i.e. a scheme which does not rely on
non-classical correlations between the atoms. 
We will assume in the following that in this case the system is subject to uncorrelated dephasing since the atoms can in principle always be put in separate setups. 
To estimate $\Omega$ or
$\delta$ with this method we use $N/2$ atoms, represented by a set
$A$, to estimate $\omega_1$ and the remaining $N/2$ atoms, represented
by a set $B$, to estimate $\omega_2$. Since the system state is a
product state these two estimations are completely independent.

Preparation of the state~(\ref{eq:pro}) is achieved by Hadamard gates
e.g. on the state $\ket{00\ldots0}$, i.e. the state of an atom $j$
before the measurement is given by
\begin{align}
  \rho_j(t) = &\frac{1}{2}\Big[ \ket{0}\bra{0} + \ket{1}\bra{1} \nonumber\\
  &+ \eta_H\e^{-\gamma t}\left( \e^{-\ii\epsilon_j t} \ket{0}\bra{1} +
    \e^{\ii\epsilon_j t} \ket{1}\bra{0}\right) \Big],
\end{align}
where $\epsilon_j= \omega_1-\omega_{L1}\equiv\varepsilon_A$ for $j\in
A$ and $\epsilon_j= \omega_2-\omega_{L2}\equiv\varepsilon_B$ for $j\in
B$. The atoms are measured in the $\sigma_x$-basis described by the
POVM~(\ref{eq:POVMX}), and therefore we obtain, e.g. for the atoms in
group $A$
\begin{eqnarray}
p(k|\omega_1) 
&=& P_+(\omega_1)^n P_-(\omega_1)^{N/2-n},
\label{eq:prob_product}
\end{eqnarray}
where 
\begin{equation}
P_\pm(\omega_1) = \frac{1}{2}\left( 1\pm\eta_H^2\eta_M\e^{-\gamma t}\cos(\varepsilon_A t) \right),
\label{eq:Ppm}
\end{equation}
and $n$ is the number of `+' measurement outcomes of the atoms in
group $A$. With the help of this we obtain 
\begin{equation}
F = \frac{1}{2}\frac{N \left(t\eta_H^2\eta_M\e^{-\gamma t}\right)^2\sin^2(\varepsilon_A t)}{1-\left( \eta_H^2\eta_M\e^{-\gamma t}\right)^2\cos^2(\varepsilon_A t)}
\end{equation}
which is maximal for $\varepsilon_A=\pi/2t$, and for $t=t_{opt}=1/2\gamma$ we have

\begin{equation}
\Delta\omega_{1,min}  = \sqrt{\frac{4\gamma\e}{NT}}\frac{1}{\eta_H^2\eta_M}.
\end{equation}
The corresponding expression for $\Delta\omega_{2,min}$ is obviously
the same and therefore we have
\begin{equation}
\Delta\Omega_{min} = \frac{\Delta\delta_{min}}{2}=\frac{\Delta\omega_{1,min}}{\sqrt 2}=\sqrt{\frac{2\gamma\e}{NT}}\frac{1}{\eta_H^2\eta_M}.
\end{equation}
The above expression has to be compared with Eq.~(\ref{eq:FI}) leading
to Eq.~(\ref{eq:fid_bound}).

\section{Maximum likelihood estimation}
\label{sec:AppIV}
To construct the Maximum likelihood estimators $\Omega_{est}$ and
$\delta_{est}$ corresponding to the DFS schemes described in 
Appendices~\ref{sec:DFS1} and~\ref{sec:DFS2}, we consider a sequence of
$\nu$ experimental runs with results $n_1,\ldots,n_\nu$, where $n_j$
is the number of times we obtain the result `+' in the $j$th repetition
of the experiment. Using $q_{n_j}(\alpha)$ from Eq.~(\ref{eq:prob}),
the likelihood function for such an outcome is given by
\begin{align}
  \mathfrak{L}(\alpha&|n_1,\ldots,n_\nu) = \prod_{j=1}^\nu \binom{N}{n_j}q_{n_j}(\alpha) \nonumber\\
  &=
  \left\{ 1+ \xi(N)\eta_H^N\eta_M^N\cos[N\varphi(\alpha)]\right\}^{\nu_e} \nonumber\\
  &\times \left\{ 1-
    \xi(N)\eta_H^N\eta_M^N\cos[N\varphi(\alpha)]\right\}^{\nu-\nu_e}
  \frac{1}{2^{\nu N}}\prod_{j=1}^\nu \binom{N}{n_j},
\label{eq:maxlik}
\end{align}
where $\varphi(\alpha)$ is given by Eq.~(\ref{eq:phialphas}), and
$\nu_e$ is the number of even $n_j$. Maximizing Eq.~(\ref{eq:maxlik})
with respect to $\alpha$ leads to the estimators
\begin{align} 
\Omega_{est} -&\frac{1}{2}(\omega_{L1}+\omega_{L2}) = \frac{\delta_{est}}{2} \nonumber\\
&=
\begin{cases}
  \frac{1}{Nt}\arccos\left(\frac{2\nu_e-\nu}{\nu\xi(N)\eta_H^N\eta_M^N}\right) \hspace{-0.2cm}&; \nu_-\le \nu_e \le \nu_+ \\
  \frac{\pi}{Nt} &;\nu_e < \nu_- \\
  0 &;\nu_e > \nu_+,
\end{cases}\nonumber\\
\end{align}
where $\nu_\pm = \nu(1\pm\xi(N)\eta_H^N\eta_M^N)/2$. The probability
distribution for $\nu_e$ is calculated to be
\begin{align}
p(\nu_e) = \binom{\nu}{\nu_e}&\left[\frac{1}{2}(1+ \xi(N)\eta_H^N\eta_M^N\cos[N\varphi(\alpha)] \right]^{\nu_e} \nonumber\\
\times &\left[\frac{1}{2}(1- \xi(N)\eta_H^N\eta_M^N\cos[N\varphi(\alpha)] \right]^{\nu-\nu_e},
\end{align}
where $\varphi(\alpha)$ is again given by Eq.~(\ref{eq:phialphas}). Using $p(\nu_e)$ we can numerically calculate the first and second moments of $\Omega_{est}$ and $\delta_{est}$ leading to the precision of estimating, e.g., $\Omega$ for finite $\nu=T/t$,
\begin{equation}
\Delta\Omega = \left\langle \left( \frac{\Omega_{est}}{\left|d\langle\Omega_{est}\rangle/d\Omega\right|}  - \Omega  \right)^2 \right\rangle^{\frac{1}{2}},
\end{equation}
which is shown in Fig.~\ref{fig3}(b).

For the sake of completeness we also give the maximum likelihood
estimator if the input state is a product state~(\ref{eq:pro}) and
undergoes uncorrelated dephasing as discussed in
Sec.~\ref{sec:AppIII}. Like before, the goal is the estimation of
$\omega_{1}$ and $\omega_{2}$ using $N/2$ atoms for each, from which we can estimate $\Omega$ and $\delta$. Since
neither the state nor the noise nor the measurement are correlated we
can treat the problem of, e.g., estimating $\omega_1$ as if there is only one atom which is measured
$\tilde\nu = N\nu/2$ times with possible outcomes $i_j=\pm$. Denoting
the total number of `+'-outcomes $n$, we obtain
\begin{equation}
  \mathfrak{L}(\omega_1|i_1,\ldots,i_{\tilde\nu}) = P_+(\omega_1)^{n} P_-(\omega_1)^{\tilde\nu-n},
\label{eq:maxlik_prod}
\end{equation}
where $P_\pm(\omega_1)$ is given by Eq.~(\ref{eq:Ppm}). Maximising $\mathfrak{L}$ leads to
\begin{eqnarray} 
\omega_{1,est} -\omega_{L1} 
&=& 
\begin{cases}
  \frac{1}{t}\arccos\left(\frac{4n-N\nu}{N\nu\eta_H^2\eta_M\e^{-\gamma t}}\right) \hspace{-0.3cm}&; \nu_-\le n\le \nu_+ \\
  \frac{\pi}{t} &;n < \nu_- \\
  0 &;n > \nu_+,
\end{cases}
\label{eq:maxlik2}
\end{eqnarray}
where $\nu_\pm = N\nu(1\pm\eta_H^2\eta_M\e^{-\gamma t})/4$.  The
probability distribution for $n$ is
\begin{equation}
p(n) = \binom{N\nu/2}{n} P_+(\omega_1)^{n} P_-(\omega_1)^{N\nu/2-n}.
\label{eq:dist}
\end{equation}
The corresponding expressions for estimating $\omega_2$ are of course
identical. The estimators for $\Omega$ and $\delta$ are then simply
given by $\Omega_{est}=(\omega_{1,est}+\omega_{2,est})/2$ and
$\delta_{est}=\omega_{1,est}-\omega_{2,est}$.  Using
Eqs.~(\ref{eq:maxlik2}) and~(\ref{eq:dist}) we can numerically
calculate $\Delta\Omega$ and $\Delta\delta$ for finite $\nu=T/t$. The
former is shown in Fig.~\ref{fig3}(b).


\end{document}